# Effect Of Coulomb Interactions on The Amplitude Of Persistent Currents In One Dimensional Disordered Mesoscopic Metallic Rings

Okanigbuan R.O[1*], Onaiwu K.N.[2] and Ehika S[1**]

[1]Department of Physics, Ambrose Alli University, PMB 14 Ekpoma, Nigeria; [2]Department of Physical and Earth Sciences, Crawford University, PMB 2001, Igbesa, Nigeria.

Email: [*]okanigban@yahoo.com, [**]ehikasimon@yahoo.com, [2]onaiwu.kingsley@crawforduniversity.edu.ng

**ABSTRACT**

Persistent current is a small but perpetual electric current that flows in metallic rings in the absence of any applied source. We compute the persistent currents of one-dimensional disordered metallic rings of interacting electrons in the presence of impurity on lattices up to 8-sites at half-filling and also away from half-filling using the Lanczos algorithm. For the case of half-filling, we observe that both interaction and disorder suppress the amplitude of the persistent currents by localizing the electrons. However, in the presence of disorder and away from half-filling, the Coulomb interaction is observed to enhance the persistent current. Furthermore, in the half-filled case, there is a transition from metal to insulator as *U* is increased significantly. In addition, shifting away from half-filling, the system is observed to remain in the metallic state irrespective of the value of the Coulomb repulsion (*U*). The observations are quite in agreement with the results from other techniques.

## 1.1 INTRODUCTION

The occurrence of persistent currents in mesocopic metallic rings threaded by a magnetic flux was first discussed by Byers and Yang (1961). After two decades later, Buttiker et al. (1983) suggested that an equilibrium persistent current could exist in a mesoscopic normal ring pierced by magnetic flux line if the size of the ring were so small that the coherence of the electrons were maintained in the entire system. This prediction was later confirmed by three experiments. The first experiment was performed on many isolated copper rings (Levy et al., 1990).

The second experiment was on a single Gold ring (Chandrasekhar et al., 1991) and the third one on a $GaAs/Al_xGa_{1-x}As$ semiconductor (Mailly et al., 1993). A long standing problem is the disagreement between theory and experiment concerning the amplitudes of persistent current in mesocopic metallic rings. The value obtained in the first experiment for the amplitude of the average persistent current is many orders of magnitude larger than that calculated by Cheung et al. (1989). The values measured in the second experiment are close to the theoretical limit for clean rings (Kato and Yoshioka, 1994).

In the third experiment, the measured current is close to the clean ring value, as is theoretically expected in this case. So many theoretical attempts have been made to explain the experimental results and the general belief is that disorder and electron-electron interaction are the key factors that can help resolve the controversy between theory and experiment. The explanation of the experimental results even becomes much more difficult since additional currents may be generated in the ring by other mechanisms which are not experimentally distinguishable from persistent currents (Kravstove and Altshuler, 2000).

A series of attempts have been made to account for this embarrassing discrepancy by the inclusion of Coulomb repulsion between electrons. In fact the main piece of evidence for the importance of this interaction was given by Schmidt (1991), where it was shown that an artificial condition of constraint which suppressed fluctuations of the electron density would strongly enhance the impurity-suppressed current. The calculation done by Ambegaoka and Eckern (1990) by including the screened Coulomb interaction via the Hartre-Forck approximation is comparable to the experimental result obtained by Levy et al.(1990). The result obtained by Eckern and Schmidt (1992), using a combined effect of impurity and Coulomb perturbation theory is comparable with experimental data. In the work done by Gambetti et al. (2002), it is shown that disorder may have the unexpected effect of favoring the zero temperature persistent current in the case of strongly interacting one-dimensional half-filled Hubbard rings. In particular for spinless fermions, it has been numerically shown that the repulsive interaction suppresses the persistent current provided that the disorder is not too strong.

The inclusion of higher order hopping integral in the tight binding Hamiltonian to study persistent current and low-field magnetic susceptibility in single isolated normal metal mesoscopic rings and cylinders (Maiti et al., 2006) gives an order magnitude enhancement of persistent current.

A calculation done using the Bethe–ansatz method (Wei et al., 2008) shows that the persistent current is suppressed by the on-site Coulomb interaction more significantly at half-filling than away from half filling. Thus in both analytical and numerical calculations the role of the electron – electron interaction and disorder on persistent current is controversial, and remain an open issue for discuss. This is the motivation for the present study.

The aim of this paper is to determine the effect of coulomb interaction and disorder on persistent currents in one-dimensional (ID) mesoscopic rings using the standard Lanczos algorithm.

We consider a system of $n$ fermions on a disordered one–dimensional lattice with Coulomb interaction. This system can be described by the Hubbard-Anderson Hamiltonian (Hubbard, 1963; Anderson, 1958)

$$H = -t \sum_{<i,j>\sigma} \left( e^{-2\pi i\phi} C^{\dagger}_{i,\sigma} C_{j,\sigma} + h.c. \right) + U \sum_i n_{i\uparrow} n_{i\downarrow} + W \sum_{i\sigma} \epsilon_i n_{i\sigma} \tag{1.1}$$

In (1.1), the energy term, $t$ is the kinetic energy term that describes the hopping of electrons between neighbouring sites through the creation operator $C^{\dagger}_{i,\sigma}$ and the annihilation operator, $C_{i,\sigma}$. The hermitian conjugate is denoted by $h.c.$, while $U$ is the on-site Coulomb repulsion that is associated with the onsite electrons through the number operator, $n_{i\sigma} = C^{\dagger}_{i,\sigma} C_{i,\sigma}$. The spin of the electron is denoted by $\sigma$, while the symbol $<i,j>$ implies that hopping is restricted to nearest neighbour sites. The symbol $W$ represents the degree of disorder or impurity strength in unit of hopping integral $t$, while symbol $\epsilon_i \in \begin{bmatrix} -1 & 1 \\ 2 & 2 \end{bmatrix}$, denotes on-sites energies. The quantum of magnetic flux $\phi$, is expressed in units of $\varphi_0 = \frac{h}{e} \approx 4.14 \times 10^{-13} \mathrm{Tm}^2$.

**1.2 METHODOLOGY**

We begin by briefly describing the standard Lanczos algorithm (Dagotto, 1994). The Lanczos method consists of the construction of a tridiagonal matrix by iteratively applying the Hamiltonian on an initial random vector say $\phi_0$ in the Hilbert space of the system under consideration. If $\phi_0$ has a non-zero projection over the true ground state $\phi_0$ of the Hamiltonian $H\phi_0 = E_0 \phi_0$, the method gives a good approximation to the ground state properties of $H$, otherwise it will converge to an excited state.

The Lanczos basis is generated using the recursion relation:

$$|\phi_{n+1}\rangle = H|\phi_n\rangle - a_n|\phi_n\rangle - b_n^2|\phi_{n-1}\rangle \qquad (1.2)$$

where

$$a_n = \frac{\langle\phi_n|H|\phi_n\rangle}{\langle\phi_n|\phi_n\rangle} \qquad (1.3)$$

and

$$b_n^2 = \frac{\langle\phi_n|\phi_n\rangle}{\langle\phi_{n+1}|\phi_{n+1}\rangle} \qquad (1.4)$$

With $b_0 = 0$ and $|\phi_{n-1}\rangle = 0$. Using the coefficients $a_n$ and $b_n$ we generate a tridiagonal matrix of the form:

$$H = \begin{bmatrix} a_0 & b_1 & & & & \cdots \\ b_1 & a_1 & b_2 & 0 & . & . \\ & b_2 & a_2 & . & . & . \\ & 0 & . & . & \ddots & b_n \\ \vdots & & & \cdots & b_n & a_n \end{bmatrix}. \qquad (1.5)$$

The persistent current is given by:

$$I_n = -\frac{1}{2\pi}\frac{\partial E_n}{\partial \phi}, \qquad (1.6)$$

where $E_a$ is calculated by exact diagonalization of the Hamiltonian. Another important parameter to be computed is the Drude weight, $D$. It is given by

$$D = \frac{N}{4\pi^2}\left(\frac{\partial^2 E_g}{\partial \phi^2}\right)_{\phi=\phi_m} \qquad (1.7)$$

It is a relevant parameter that characterizes the insulating and conducting properties of the system. If

$D_c = 0$, the material is insulating, if $D_c > 0$, the material has infinite direct current conductivity and is a perfect metal.

In the insulating regime, the amplitude of the flux dependence oscillation of the ground state energy is much smaller than the energy gap between the many body ground state and the first excited state. Therefore perturbation theory in the hopping matrix elements across the boundary is enough to describe the flux dependence of the ground state energy (Weinmann et al, 2001). This ground state is given by

$$E(\phi) = E(0) - \frac{\Delta E}{2}[1 - \cos(\Phi)] \qquad (1.8)$$

where $\Delta E = E(0) - E(\pi)$ can be interpreted as the difference of ground state energies between periodic $(\Phi = 0)$ and anti-periodic $(\Phi = 2\pi\phi)$ boundary conditions, since a magnetic flux through the ring is equivalent to introducing a change of the boundary conditions.

The persistent current $I(\phi)$ is given by:

$$I(\phi) = -\frac{dE(\phi)}{d\phi} = \frac{\Delta E}{2\phi_0}\sin(2\pi\phi). \qquad (1.9)$$

Similarly, the Drude weight can also be expressed in terms of $\Delta E$ as

$$D_c = -\frac{N}{2}\Delta E \qquad (1.10)$$

To give a brief illustration of our method, we solve a simple two-site problem of two interacting electrons. The two-site chain in the subspace $S_z = 0$ is defined by the basis:

$|1\rangle = |1\uparrow, 1\downarrow\rangle; |2\rangle = |2\uparrow, 2\downarrow\rangle; |3\rangle = |1\uparrow, 2\downarrow\rangle; |4\rangle = |1\downarrow, 2\uparrow\rangle.$

We randomly choose our initial vector to be $|3\rangle$, that is $|3\rangle = |\phi_0\rangle = |1, 2 \downarrow\rangle$, and for simplicity we take $W = 0$ in the Hamiltonian (1.1).

Using the Lanczos procedure above, we generate the following matrix:

$$H = \begin{bmatrix} 0 & t\sqrt{2}\cos(4\pi\phi) & 0 \\ t\sqrt{2}\cos(4\pi\phi) & U & \dfrac{t\sqrt{4(1-\cos(4\pi\phi))+2\cos^2(4\pi\phi)}}{\sqrt{\cos(4\pi\phi)}} \\ 0 & \dfrac{t\sqrt{4(1-\cos(4\pi\phi))+2\cos^2(4\pi\phi)}}{\sqrt{\cos(4\pi\phi)}} & 0 \end{bmatrix}. \quad (1.11)$$

It is easy to show that the ground state energy of (1.11) is

$$E_g = \frac{U}{2} - \frac{1}{2}\left[U^2 + 16t^2\left(\cos(4\pi\phi) - 1 + \frac{1}{\cos(4\pi\phi)}\right)\right]^{\frac{1}{2}} \quad (1.12)$$

The current in the ring is obtained from (1.6) as

$$I(\phi) = \frac{8t^2\left(\dfrac{\sin(4\pi\phi)}{\cos^2(4\pi\phi)} - \sin(4\pi\phi)\right)}{\left[U^2 - 16t^2 + 16t^2\cos(4\pi\phi) + \dfrac{16t^2}{\cos(4\pi\phi)}\right]^{\frac{1}{2}}} \quad (1.13)$$

and the Drude weight is obtained as

$$D_c = \left[\frac{\left(\dfrac{64t^2\sin(4\pi\phi)}{\cos^2(4\pi\phi)} - 64t^2\sin(4\pi\phi)\right)^2}{16\left(16\cos^2(4\pi\phi) + \dfrac{16t^2}{\cos(4\pi\phi)} + 16t^2 + U^2\right)^{3/2}} - \frac{\dfrac{512\sin^2(4\pi\phi)t^2}{\cos^3(4\pi\phi)} - 256t^2\cos(4\pi\phi) + \dfrac{256t^2}{\cos(4\pi\phi)}}{8\sqrt{16t^2\cos(4\pi\phi) + \dfrac{16t^2}{\cos(4\pi\phi)} - 16t^2 + U^2}}\right]_{\phi=0} \quad (1.14)$$

or $D_c = 0$.

**1.3 RESULTS AND DISCUSSION**

In Fig. 1, Persistent current is plotted against the magnetic flux $\phi$ for a 2-site ring for ordered interacting case $W = 0$ using (1.13). The current amplitude is suppressed as the repulsive Coulomb is increased from $U = 0$ up to $U = 8$. The Coulomb repulsion suppresses the current amplitude by localizing the electrons. The current is found to oscillate with the magnetic flux, with a period $\phi_0/2$.

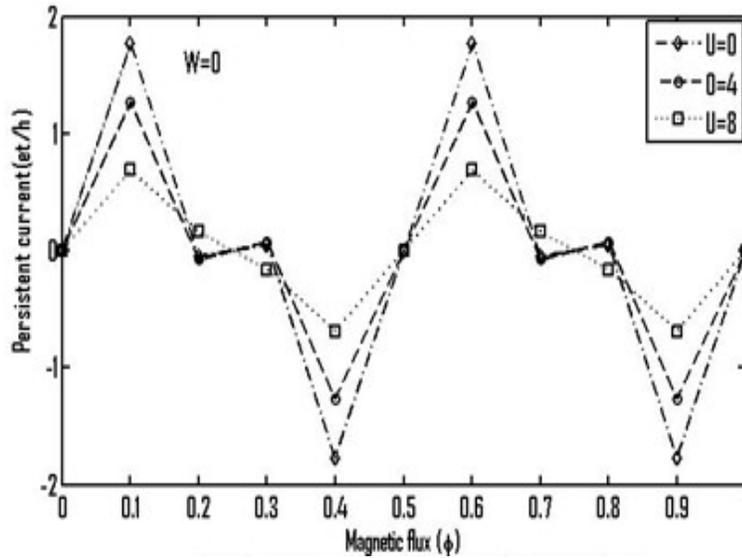

*Figure 1: Persistent current vs magnetic flux for 2 electrons on 2 sites*

The influence of repulsive interactions at a fixed disorder strength $W = 1$ for the case of half-filing $N = n = 8$, off half-filling $N = 8, n = 6$ and off half-filling $N = 8, n = 2$ are respectively shown in Figs. 2, 4 and 7. At half filling the current amplitudes is strongly suppressed by both the coulomb repulsion and impurity (disorder). Away from half filling, the effect of interactions are much weaker, that is the current amplitudes is less affected by moderate interactions. Our results are comparable with those obtained by Wei et al., (2008) who worked on half-filled and off half-filled 66 sites one-dimensional ring in the absence of impurity. It also compares well with the result obtained by Bouzerar et al., (1994) for a 16 sites ring.

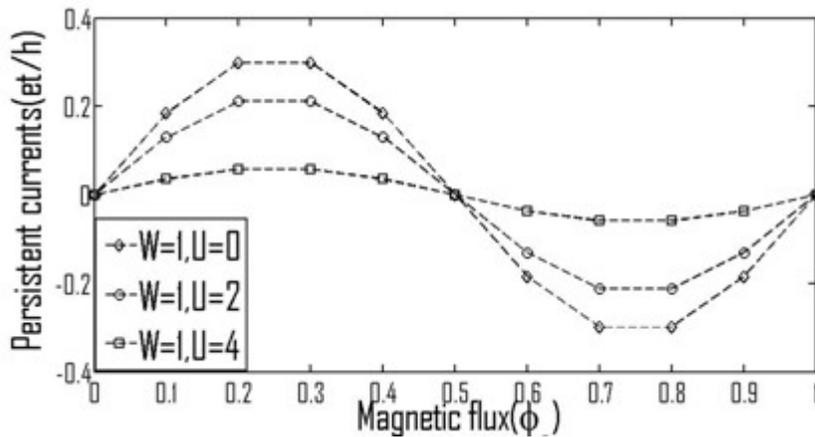

*Figure 2: Variation of the persistent currents with magnetic flux for 8 electrons on 8-sites*

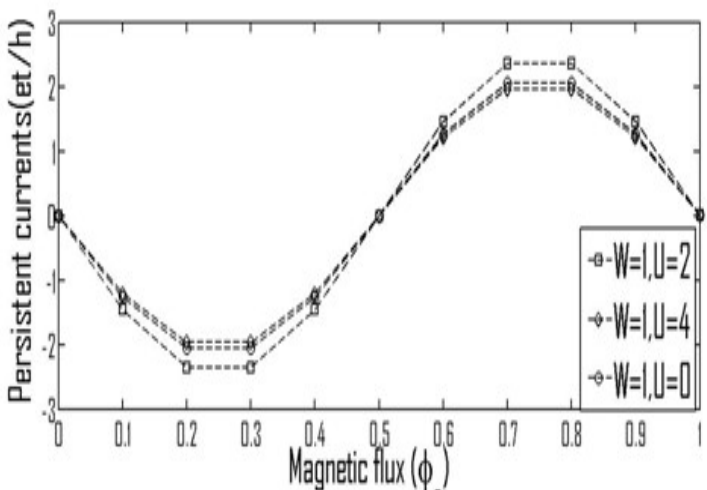

*Figure 3: Persistent currents versus Magnetic flux for 6 electrons on 8-sites.*

Of particular interest is Fig. 4 where the current is enhanced by a factor of 1.24 when the coulomb repulsion is increased from U=0 to U=2. This is because electron interaction has the effect of of kicking out electrons from their preferred site, thus delocalizing them, obviously increasing the persistent current.

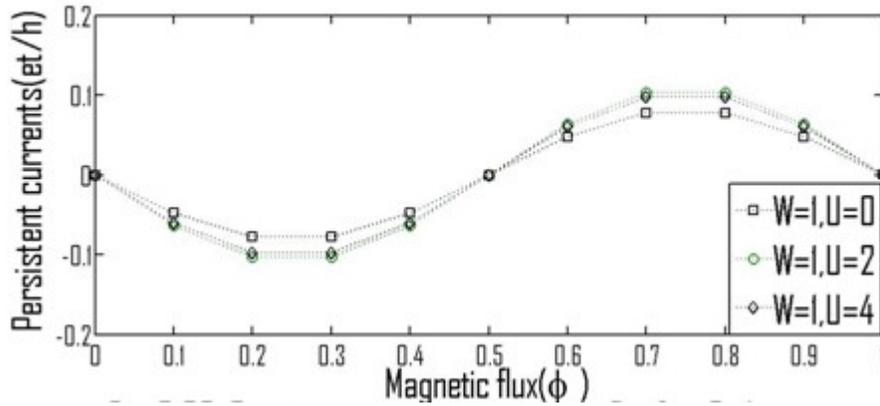

*Figure 4: Persistent currents versus magnetic flux for 2 electrons of 8-sites.*

Thus electron-electron interactions tend to homogenize the system, offsetting the reduction in the current promoted by disorder.

In order to characterize the transport properties of the system under study, the Drude weight as a function of coulomb interaction (U) is computed for half filling and away from half filling.

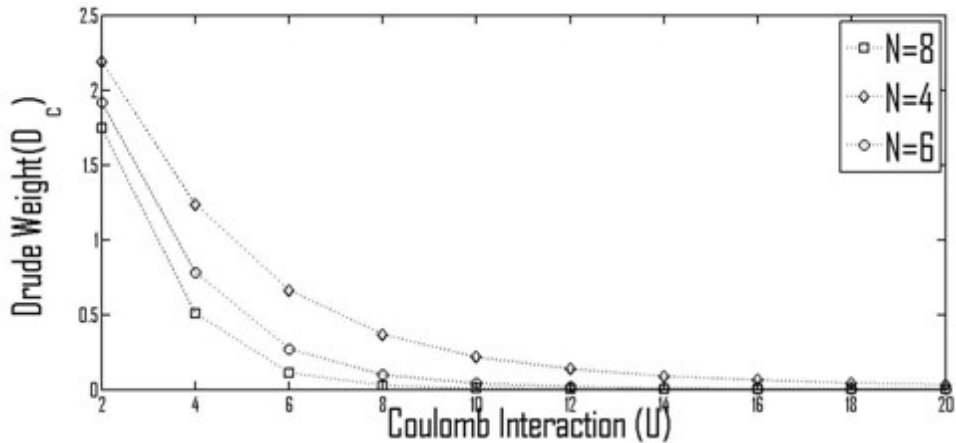

*Figure 5: Drude Weight versus Coulomb interaction at half filling*

Firstly we consider a system of two interacting electrons system where the Drude weight vanishes for all values of Coulomb repulsion (U). That is $D_c = 0$, as indicated in (1.14). This implies that the 2-electron system is an insulator. There is a transition from metallic phase, when $D_C$ is finite, to an insulating phase, when the $D_C$ vanishes (fig. 5). This is a manifestation of the Mott-Hubbard metal-insulator transition and is comparable to the result obtained by Maiti (2010).

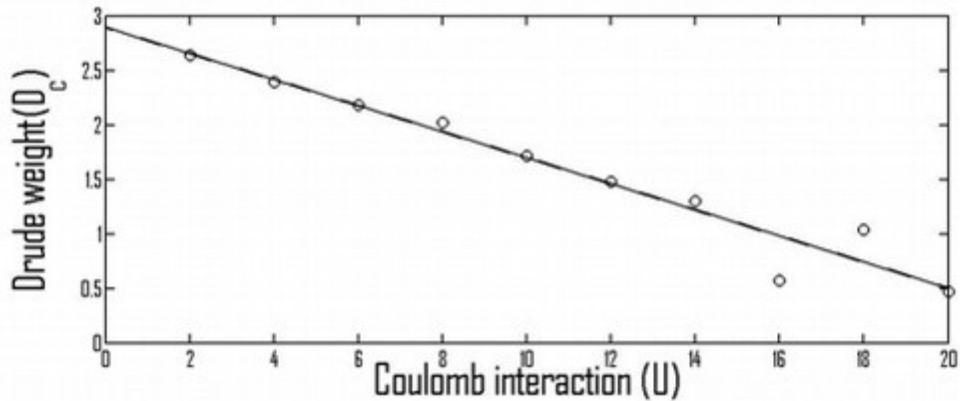

*Figure 6: Drude weight VS Coulomb interactions for four electrons on six sites of ID Hubbard model*

Away from half filling, (fig. 6), the ring is always conducting irrespective of the value of the Coulomb repulsion $U$. This is comparable to the result obtained by Akande and Sanvito (2012) for 30 electrons on 60-sites using the current density functional theory (CDFT). The results in fig. 5 and fig. 6 also compare nicely with the result obtained by Maiti et al., (2004), Saha and Maiti (2016) for 1-D Hubbard rings at half-filled rings and non-half filled rings respectively.

## 1.4 CONCLUSIONS

This study looked at the persistent currents of a one-dimensional disordered metallic rings using the Lanczos algorithm. It was found that at half-filling, both Coulomb interaction and disorder strongly suppress the persistent current by localizing the electrons. In addition, at half filling, the Coulomb interaction induces a metal-insulator transition. However, away from half-filling the system remains in the metallic phase irrespective of the value of the Coulomb repulsion. But off half-filling, it was observed that, there is an increase of current amplitude due to Coulomb interaction in the presence of impurity.

ACKNOWLEDGEMENTS

Okanigbuan R.O. is thankful to the International Centre for Theoretical Physics(ICTP) Trieste, Italy, where I did some of the computation on the work. Special thanks go to Prof J.O.A. Idiodi, Prof. Vladimir E. Kravstove, Prof Muller and Hon-Yi Xie for stimulating discussions.

**REFERENCES**

Akande A. and Sanvito S., J. Phys. Condens. Matter **24,** 1-12, (2012).
Ambegaokar V. and Eckern U., Phys. Rev. Lett. **65**, 381-384, (1990).
Anderson P. W., Phys. Rev. **109,** 1492-1505, (1958).
Bouzerar G., Poilblank D., and Montambaux G., Phys. Rev. B **49**, 8258-8262, (1994).
Buttiker M., Imry Y., and Landauer R., Phys. Lett. **96,** A, 365-367, (1983).

Byers N. and Yang N., Phys. Rev. Lett. **7**, 46-49, (1961).
Chandrasekhar V., Webb R.A., Brady J., kitchen M.B., Gallagher W.J., and Kleinassar A., Phys. Rev. Lett. **67**, 3578-3581, (1991).
Cheung F., Riedel K., and Gefen Y., Phys. Rev. Lett. **62**, 587-590, (1989).
Dagotto E., Rev. Mod. Phys. **66,** 763, (1994).
Eckern U. and Schmid A., Europhys. Lett. **18,** 457-462, (1992).
Gambetti E., Phys. Rev. B **72**, 165338-1-15, (2005).
Gambetti E., Weinmann R., Jallabat R., and Brune P. H., Europhys. Lett., **60**, 120-126, (2002).
Hubbard J., Proc. R. Soc. **276,** 1365, (1963).
Karmakar N.S., Maiti S. K. and Chowdhury J., Springer series in Solid-State Sciences: 232, (2007).
Kato H. and Yoshioka D. (1994). Phys. Rev. B **50,** 4943-4946.
Kravtsov V. E. and Altshuler B. L., Phys. Rev. Lett. **84**, 3394-3397, (2000).
Levy L. P., Dolan G., Dunsmuir J. and Bouchiat. Phys. Rev. Lett. **64**, 2074-2077, (1990).
Mailly D., Chapelier C.and Benoit A., Phys. Rev. Lett. **70**, 2020-2023, (1993).
Maiti S. K. and Saha M., Phys. Lett. A **380**, 1450-1454, (2016).
Maiti S. K., Solid statecommun. **150**, 2212-2217, (2010).
Maiti S. K., Chowdbury J., and Karmakar S. N., Phys. Lett. A332, 497-502, (2004).
Schmid A., Phys. Rev. Lett. **66**, 80-83, **(**1991).
Wei Bo-Bo, Shi-Jian G., and Hai-Qing L., J. Phys. Condens. Matter **20**, 1-8, (2008).
Weinmann D., Schmitteckert P., Jalabert R. A. and Pichard J. L., Eur. Phy. J. B **19**, 139-156, (2001).
Yoshioka D., and Kato H., Physica B **212**, 255, (1995).